 \documentclass[pra,twocolumn]{revtex4}

 \usepackage{graphicx}

\begin{document}

 \def\BE{\begin{equation}}
 \def\EE{\end{equation}}
 \def\BA{\begin{array}}
 \def\EA{\end{array}}
 \def\BEA{\begin{eqnarray}}
 \def\EEA{\end{eqnarray}}
 \def\nn{\nonumber}
 \def\ra{\rangle}
 \def\la{\langle}

 \title{Schr{\"o}dinger cat state preparation by non-Gaussian continuous variable gate }
 \author{Ivan~V.~Sokolov}
 \email{i.sokolov@mail.spbu.ru}
 \affiliation{Saint Petersburg State University,  7/9 Universitetskaya nab., 199034  Saint Petersburg, Russia}

 \begin{abstract}

We propose a non-Gaussian continuous variable (CV)  gate which is able to conditionally produce superposition
of two ``copies'' of an arbitrary input state well separated in the coordinate and momentum plane --
 a Schr{\"o}dinger cate state. The gate uses cubic phase state of an ancillary oscillator as a
non-Gaussian resource, an entangling Gaussian gate, and homodyne measurement which provides
nonunique information about the target system canonical variables, which is a key feature of the scheme.
We show that this nonuniqueness manifests problems which may arise by extension of
the Heisenberg picture onto the measurement-induced evolution of CV non-Gaussian networks,
if this is done in an approach commonly used for  CV Gaussian schemes of quantum information.

 \end{abstract}


 \maketitle


Continuous variable  quantum networks are a promising area of quantum information (quantum communication,
quantum computing and simulation)  \cite{Lloyd99,Braunstein05}. Since the pioneering demonstrations of  CV quantum
teleportation \cite{Furusawa98} and quantum dense coding \cite{Li02}, there was achieved an impressive progress
in the theory and implementation of multimode CV cluster schemes \cite{Gu09,Weedbrook12,Ukai15}.
Such schemes are based on Gaussian operations: the linear and bilinear in canonical variables interactions,
homodyne measurements and feedforward. The large-scale CV cluster states multiplexed in the time or wavelength domain
were realized and characterized \cite{Yokoyama13,Roslund14}.

The CV Gaussian quantum evolution can be effectively simulated using classical computer.
In order to achieve quantum supremacy, additional non-Gaussian operations are needed \cite{Lloyd99,Bartlett02}. This can be
done in different approaches, including hybrid discrete- and continuous variable schemes \cite{Andersen15,Ra20}
or by making use of non-Gaussian Hamiltonians of the order higher than two. A cubic phase gate
\cite{Gottesman01,Bartlett_Sanders02}  based on cubic non-demolition single-mode interaction and
homodyne measurement, or  higher order phase gates can serve as building blocks for universal quantum computing.
Any given exponential operator of bosonic field operators, describing an arbitrary multimode Hamiltonian evolution,
can be systematically decomposed into a set of universal unitary gates \cite{Sefi11}.

An implementation of cubic (or higher order) phase state meets difficulties due to weak non-linear coupling of the
interacting bosons in most systems. To get around this problem, it was initially proposed  \cite{Gottesman01}
to prepare two quadrature entangled oscillators and to perform the ancillary oscillator homodyne measurement
retaining non-linear terms which are usually omitted (see also \cite{Ghose07}).
An approximative weak cubic state described as a superposition
of first low-photon Fock states was proposed \cite{Marek11} and experimentally prepared \cite{Yukawa13}.
An approach to generating the cubic phase gate was introduced \cite{Marshall15} where a target oscillator is repeatedly
entangled with a weak coherent ancilla, and the ancilla photon subtraction or projection measurement is performed.

The general properties of the CV schemes with embedded non-Gaussian elements are actively explored today.
An adaptive non-Gaussian measurement can be used in order to implement the cubic phase gate \cite{Miyata16},
and an approach was demonstrated \cite{Marek18} that allows effectively merge a sequence of single-mode
non-Gaussian gates in order to achieve a given operation.
The CV hypergraph cluster states with three-mode cubic nonlinearity
can be used for universal quantum computing \cite{Moore19}.

In this work we present a CV non-Gaussian gate which conditionally generates a Shr{\"o}dinger cat state from an
arbitrary target state that occupies a finite area on the phase plane. The gate exploits the same elements as the
cubic phase gate introduced in \cite{Gottesman01,Bartlett_Sanders02}, that is, the cubic phase state, a two-mode
entangling Gaussian operation, and  homodyne measurement of an ancillary oscillator.
As an output, a superposition of two copies of the target states well separated along the momentum  axis
(or, in general, along an arbitrary direction) emerge. By sequentially applying operations of this kind together
with standard Gaussian gates that generate shift, rotation, squeezing, shearing, etc.,
one can transform an initial CV network quantum state to a Shr{\"o}dinger cat state of an arbitrary complexity.

The Schr{\"o}dinger cat states are an object of a relentless interest since their first introduction \cite{Schrodinger35}.
Besides their fundamental importance, some proposals for fault-tolerant quantum information processing directly rely
on the cat-like states \cite{Gottesman01,Mirrahimi14}.

In general, one can prepare CV Schr{\"o}dinger cat states by making use of the unitary evolution assisted by a
non-linear interaction, as for example in the earlier proposal \cite{Yurke86}, where a huge Kerr nonlinearity is involved.
The schemes based on a measurement-induced evolution also can create cat-like states.
The optical squeezed Schr{\"o}dinger cat states were generated in low-photon regime using homodyne
detection and photon number states as resources \cite{Ourjoumtsev07}.
An iterative scheme where  the target state is subsequently  mixed on a passive
beamsplitter with a heralded single-photon ancillary state and homodyne measurement is performed, was
discussed  and implemented experimentally \cite{Etesse14,Etesse15}. The proposals
\cite{Ourjoumtsev07,Etesse14,Etesse15} are specifically aimed at the creation of superpositions of two coherent states.

Unlike the previously introduced CV schemes with  non-Gaussian gates, we consider  measurement  of the
ancillary oscillator that provides  nonunique information about the relevant physical variables of the target subsystem,
which is a key point. This feature does not emerge in Gaussian quantum networks and manifests a superior complexity
of the non-Gaussian schemes. The approach used in our proposal is scalable, that is, it can be extended to more
complex CV non-Gaussian gates. From an heuristic point of view, one can easily identify some confi\-gu\-ra\-tions
where the CV cat states may arise using illustrative schemes similar to the one presented below.

 \begin{figure}
 \begin{center}
 \includegraphics[width=0.65\linewidth]{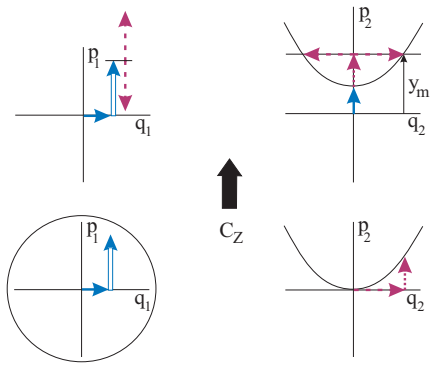}
 \caption{Measurement-induced evolution of the position $q_i$ and momentum $p_i$ of the target
field (left) and the ancillary oscillator (right), where the ancilla is initially prepared in cubic phase state.
Here different arrows schematically represent quadrature
amplitudes $\{q_i,p_i\}$ randomly chosen within the uncertainty region of the corresponding field.
The ancilla wave function support is approximated with parabola, see comments in the text.
The Gaussian $C_Z$ operation entangles two oscillators.
A Schr{\"o}dinger cat state emerge when the
ancilla momentum measurement outcome $p_2\to y_m$ is compatible not with one, but with two clearly
distinguishable values of the ancilla coordinate and, due to the entanglement, of the output field momentum.
In quantum description this measurement-induced nonuniqueness may result in the output field
superposition state whose components are mutually separated on the phase plane.}
 \label{figure_1}
 \end{center}
 \end{figure}

The target and ancillary oscillators evolution is schematically shown in Fig.~\ref{figure_1} which in graph form illustrates
transformations of the relevant canonical variables. The ancilla is assumed to be initially prepared in the cubic phase state
\cite{Gottesman01,Bartlett_Sanders02} via the action of unitary evolution operator $\exp(i\gamma q_2^3)$ upon the
momentum eigenstate $|0\ra_{p_2}$. In the Heisenberg picture this may be written as
 \BE
 \label{cubic_Heisenberg}
 q_2 = q_2^{(0)}, \qquad p_2 = p_2^{(0)} + 3\gamma q_2^2,
 \EE
where $p_2^{(0)}|0\ra_{p_2}=0$.

The two-mode entangling operation $C_Z\sim\exp(iq_1q_2)$ transforms the system variables as
 \BE
 \label{C_Z}
\BA{ll}q'_1=q_1\\p'_1=p_1+q_2,\EA
\qquad \BA{ll}q'_2=q_2\\p'_2 = p_2+q_1.\EA
 \EE
In order to illustrate the system evolution before the measurement, we represent initial state of both
oscillators as a statistical ensemble of canonical coordinates and momenta $\{q_i,p_i\}$ in a semiclassical manner.
Since the uncertainty region of the state $|0\ra_{p_2}$ is horizontal straight line, by applying (\ref{cubic_Heisenberg})
one arrives to an approximate representation of the cubic phase state support in the form of a parabola.

In general, the cubic phase state Wigner function \cite{Ghose07} has fringes and negative
values in some areas on the phase plain, which of course is missing in the simplified picture.
Nevertheless, in the following we demonstrate that this illustrative approach provides a hint allowing to infer
the cat state emergence, and even to deduce its properties under certain limitations.

In Fig.~\ref{figure_1} quadrature amplitudes are randomly chosen on the phase plane within the uncertainty
region of the corresponding field and represented by the arrows. Next, the  transformation (\ref{C_Z}) is applied.

In the Schr{\"o}dinger picture,  the initially prepared independent input field and ancilla  states are
 $$
|\psi_1\ra = \int dx_1\psi(x_1)|x_1\ra,
 $$
 $$
|\psi_2\ra = e^{i\gamma q_2^3}\int dx_2|x_2\ra = \int dx_2e^{i\gamma x_2^3}|x_2\ra,
 $$
correspondingly, where the input field coordinate wave function is $\psi(x_1)$.  The cubic phase state
of the ancillary oscillator  emerges during non-Gaussian evolution of the unnormalizable momentum eigenstate
$|0\ra_{p_2}$. The two-mode entangling operation $\exp(iq_1q_2)$ acts on the initial state with the outcome
 \BE
 \label{after_Cz}
 |\psi_{12}\ra = \int dx_1dx_2\psi(x_1)e^{ix_2(x_1 + \gamma x_2^2)}|x_1\ra|x_2\ra.
 \EE
In order to describe measurement-induced state reduction, we represent the state (\ref{after_Cz}) in the measurement
basis, using standard definitions
 $$
|x\ra = \frac{1}{\sqrt{2\pi}}\int dy e^{-ixy}|y\ra, \quad |y\ra = \frac{1}{\sqrt{2\pi}}\int dx e^{ixy}|x\ra.
 $$
In the following, we use $x$ and $y$ to label the coordinate and momentum eigenstates correspondingly.
The state (\ref{after_Cz}) may be written as
 $$
\frac{1}{\sqrt{2\pi}} \int dy_2|y_2\ra\int dx_1 dx_2\,\psi(x_1)e^{ix_2(x_1 - y_2 + \gamma x_2^2)}|x_1\ra.
 $$
The ancilla momentum measurement with the outcome $y_m$ projects the first oscillator into the state with the
wave function  $\tilde\psi(x) = {\cal N} \psi(x)\varphi_\gamma(x - y_m)$. Here ${\cal N}$ is the normalization
coefficient and the added factor is expressed in terms of the Airy function,
 \BE
 \label{psi1_reduced}
\varphi_\gamma(x-y_m) =\frac{1}{\sqrt{2\pi}}\int dx'e^{ix'(x - y_m + \gamma x'^2)} =
 \EE
 $$
\Big(\sqrt{2\pi}/(3\gamma)^{1/3}\Big) {\rm Ai}\left((x - y_m)/(3\gamma)^{1/3}\right).
 $$
The output field coordinate $x$ in (\ref{psi1_reduced}) is in the range that supports the spatial
distribution $|\psi(x)|^2$. A Schr\"odinger cat state conditionally arise when the measured ancilla momentum is large
enough, $y_m > x$ for all $x$ within this range.

For such measurement outcome, the polynomial $x'(x - y_m + \gamma x'^2)$ in the exponent can be approximated with
quadratic expression. In Fig.~\ref{figure_1} (upper right corner) this corresponds to the limit where the curve that
approximately represents the displaced cubic phase state might be replaced with straight lines in a vicinity of two
intersections with the horizontal line indicating the measured momentum.

The stationary phase points are $x_s' = \pm\sqrt{(\gamma_m - x)/3\gamma}$, and the exponent
in (\ref{psi1_reduced}) is represented near these points as
 $$
\mp \frac{2}{3\sqrt{3\gamma}}(y_m - x)^{3/2} \pm
\sqrt{3\gamma(y_m - x)}\left(x' \mp \sqrt{(y_m-x)/3\gamma}\right)^2.
 $$
The factor (\ref{psi1_reduced}) becomes a sum of two independent integrals. By making use of
 $$
 \int dx e^{i\kappa x^2} = \exp\left(i\frac{\pi}{4}\right) \sqrt{\frac{\pi}{\kappa}},
 $$
we finally identify the output field state wave function as a superposition of two distinguishable components,
 \BE
 \label{cat_approx}
 \tilde\psi(x) = {\cal N}\psi(x)\left(\varphi_\gamma^{(+)}(x - y_m) + c.\, c.\right),
 \EE
where
 \BE
 \label{factor_approx}
\varphi_\gamma^{(+)}(x - y_m) =
 \EE
 $$
\exp\left[i\left(\frac{\pi}{4} - \frac{2}{3\sqrt{3\gamma}}(y_m - x)^{3/2}\right)\right]
\big(12\gamma(y_m - x)\big)^{-1/4}.
 $$
For $y_m \gg |x|$ within the relevant range, the exponent can be linearized in $x/\gamma_m$.
This yields $\varphi_\gamma^{(+)}(x - y_m)\sim \exp\big(i\sqrt{y_m/3\gamma}\, x\big)$,
which means that this gate conditionally projects the initial state into the superposition of two ``copies'' with
well-defined opposite shift in momentum of $\pm\sqrt{y_m/3\gamma}$, in agreement with Fig.~\ref{figure_1}
(upper left corner). If this measurement-dependent shift is larger than the range of momentum
spanned by the input state, the gate output state is nothing but a Schr\"odinger cat.

 \begin{figure}
 \begin{center}
 \includegraphics[width=0.65\linewidth]{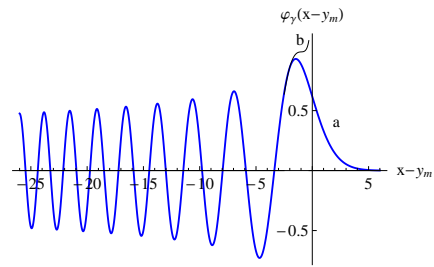}
 \caption{The factor $\varphi_\gamma(x-y_m)$  added by the gate to the initial coordinate wave
function by the outcome $y_m$ of ancilla momentum measurement.  Here $\gamma=1$, curve $a$
was evaluated numerically from the representation of the Airy function through the integral (\ref{psi1_reduced}),
curve $b$ corresponds to the approximate solution (\ref{cat_approx},\ref{factor_approx}).}
 \label{figure_2}
 \end{center}
 \end{figure}

The factor added by the gate is represented \footnote{In order to ensure better
convergency of the improper integral (\ref{psi1_reduced}) that represents the Airy function, one can apply deformation
of the integration contour in the complex plain by assuming $x' = ze^{i\pi/6}$ for $z>0$.} in Fig.~\ref{figure_2}.
As seen from the figure, the exact factor  $\varphi_\gamma(x - y_m)$ and the approximate one are in
perfect agreement except for a small range of coordinate. For $x \approx y_m$, the approximate solution diverges
since the approximation of two distinguishable crossings does not work near the bottom of parabola in Fig.~\ref{figure_1}.
In this area the non-classical properties of the cubic phase state Wigner function are most pronounced.
The exact result  exhibits fast decrease in the region $y_m<x$ which is unreachable for the measurement
in classical picture.

The cat-breeding transformations of an arbitrary input state may be combined with standard Gaussian
operations such as displacement, rotation, squeezing and shearing. By establishing the measurement window
with a needed precision (which of course has effect on the success probability), one can conditionally generate on demand
complex cat-like structures on the phase plane.\vspace{2mm}

There is a variety of measurement-based CV Gaussian schemes of quantum information, such as quantum teleportation
and dense coding, quantum repeaters, cluster model of quantum computing, etc.
The Gaussian schemes with measurement and feedforward can be effectively described both in the Schr\"odinger and
the Heisenberg picture, where the last one offers a possibility to include noise sources and imperfections in the
scheme  \cite{Gu09,Weedbrook12,Ukai15,Yokoyama13,Roslund14,Korolev20} and provides in many cases an intuitively
clear illustration of the processes in the device.

Various approaches towards more general configurations seeded with non-Gaussian gates were developed,
and some works analyzed such gates both in the Schr{\"o}dinger and the Heisenberg picture \cite{Miyata16,Marek18}.

The non-Gaussian gate described here exploits the same key elements (that is, a nonlinear phase state, an entangling
Gaussian gate and homodyne measurement) as many other schemes. This gate provides an instructive example where the
Heisenberg picture, when applied within the same paradigm as in the measurement based Gaussian quantum information,
fails due to the presence of non-Gaussianity.

The transformations (\ref{cubic_Heisenberg}, \ref{C_Z}) which describe the system unitary evolution in the Heisenberg picture
before the measurement may be combined as
 $$
\BA{ll}q'_1=q_1\\p'_1=p_1+q_2,\EA
\qquad \BA{ll}q'_2=q_2\\p'_2 = p_2+q_1=p^{(0)}_2 + 3\gamma q_2^2+q_1.\EA
 $$
Following standard for Gaussian networks approach, we describe the ancilla momentum measurement  by
substituting the measurement outcome $y_m$ for the operator-valued amplitude $p'_2$. This
yields  $3\gamma q_2^2 = y_m - q_1 - p_2^{(0)}$.
Then $q_2$ is excluded from  $p'_1$, and we arrive at the target oscillator output quadrature amplitudes
 \BE
 \label{cubic_cat_transf}
 \BA{ll}q^{(out)}_1=q_1\\p^{(out)}_1 = p_1 \pm (3\gamma)^{-1/2}\sqrt{y_m - q_1 -p_2^{(0)}}.\EA
 \EE
Since statistical averaging of the canonical variables and their momenta in the Heisenberg picture must be performed
with the system initial state and in view of  $p_2^{(0)}|0\ra_{p_2}=0$, we drop $p_2^{(0)}$ here.

There are evident problems with this result which make it useless for the standard task: to evaluate correlation
functions and momenta  of the output physical quantities. First, the expression for $p_1^{(out)}$ is  nonunique.
Secondly, as far as the factor $\varphi_\gamma(x-y_m)$ is not equal zero beyond the classically unreachable region,
the measurement outcomes $y_m<x$ are possible which makes $p_1^{(out)}$ look non-Hermitian.

On the other hand, a comparison with the results derived above from the Schr\"odinger
picture makes it clear that just this nonuniqueness manifests emergence of a Schr\"odinger cat state.

It is worth noting that even when the output physical quantities emerge as single valued, this does not guarantee correct
description of a CV non-Gaussian scheme with measurement and feedforward, if the Heisenberg picture is used within
the same paradigm as in Gaussian quantum information, as we shall discuss elsewhere.\vspace{2mm}

In conclusion, we have presented non-Gaussian CV gate which is able to conditionally produce a superposition
of two ``copies'' of an arbitrary input state, well separated in the coordinate and momentum plane, which corresponds
to the  emergence of Schr{\"o}dinger cat state.
In analogy to some other proposals, the scheme exploits the same key elements, that is,
ancillary cubic phase state and $C_Z$ Gaussian gate, but applies homodyne measurement which provides
a nonunique information about the target system physical quantities. One can infer from this result that a CV quantum
network with embedded non-Gaussian gates of  general kind may experience measurement-induced evolution into a
Schr\"odinger cat state of an arbitrary complexity.

The scheme presented here also provides an instructive example of problems which may arise by the extension of
the Heisenberg picture, as it is commonly used in various CV Gaussian schemes of quantum information,
onto the measurement-induced evolution of CV non-Gaussian networks.\vspace{2mm}

I thank Nikolay N. Bezuglov for fruitful discussions. This research was supported by the Russian Foundation for
Basic Research (RFBR) under the projects 18-02-00648-a and 19-02-00204-a.


 \bibliography{}
 \bibliographystyle{plain}

\end{document}